\documentclass{emulateapj}

\newcommand{\be}{\begin{equation}}
\newcommand{\ee}{\end{equation}}
\newcommand{\beqn}{\begin{eqnarray}}
\newcommand{\eeqn}{\end{eqnarray}}
\newcommand{\bi}{\begin{itemize}}
\newcommand{\ei}{\end{itemize}}

\begin{document}

\title{The Eccentric Kozai Mechanism for a Test Particle}
\author{Yoram Lithwick\altaffilmark{1} and Smadar Naoz\altaffilmark{1}}
\altaffiltext{1}{Department of Physics and Astronomy, Northwestern University, 2145 Sheridan Rd., Evanston, IL 60208}

\begin{abstract}
We study the dynamical evolution of a test particle that 
orbits a star
in the presence of an exterior massive planet, considering
octupole-order secular interactions.
In the standard Kozai mechanism (SKM),
 the planet's orbit is circular, and so
the particle conserves vertical angular momentum.
As a result, the particle's 
 orbit oscillates periodically, exchanging
eccentricity for inclination.
However, when the planet's orbit is eccentric,
the particle's vertical angular momentum varies
and its Kozai oscillations are modulated on longer timescales---we
call this the eccentric Kozai mechanism (EKM).
The EKM can lead to behavior that is dramatically different from 
the SKM.  In particular, the particle's orbit can flip from prograde
to retrograde and back again, and it can reach arbitrarily high eccentricities
given enough time.   
We map out the conditions under which this dramatic
behavior (flipping and extreme eccentricities) occurs,
and show that
when the planet's eccentricity is sufficiently high, 
it
occurs quite generically.
For example, when the planet's eccentricity
exceeds a few percent of the
ratio of semi-major axes (outer to inner),  
around half of randomly oriented test particle
orbits will flip and reach extreme eccentricities.
The SKM has often been invoked for bringing pairs 
of astronomical bodies
 (star-star, planet-star, compact-object pairs)
  close together.
    Including
the effect of the EKM will enhance the rate at which 
such matchmaking occurs.

\end{abstract}

\section{Introduction}

\cite{Naoz10} recently showed
 that
in a system with two planets orbiting
a star,
the inner planet's orbit can flip from prograde
to retrograde and back, and  can also reach extremely
high eccentricities.
Therefore, 
 starting from a system with two prograde
planets on distant orbits from their star,
 the inner planet
can
both flip its orbital orientation and reach high enough eccentricities
to be tidally captured by the star.  
That might explain the origin of the $\sim 25\%$ of hot Jupiters
whose orbits are retrograde with respect to the spin
of their star, as inferred from
 Rossiter-McLaughlin measurements
  \citep[e.g.,][]{Tri+10}.
  However, while the behavior   
 displayed in \cite{Naoz10} is intriguing, it has not yet been 
understood. Furthermore,  \cite{Naoz10}
choose  initial eccentricities and inclinations that
are fairly extreme for planetary systems
($e\sim 0.6, i\sim 70^\circ$), 
and hence it is not yet clear how generic their results are.
The goal of this paper is to understand and map out the mechanism
in a simplified system where the inner planet is treated 
as a massless test particle.

There has been much work in the literature
on the orbital dynamics of similar systems.
\cite{Kozai}
considered the evolution of an asteroid perturbed by a circular Jupiter.
He focused on secular interactions, meaning
that interactions are averaged over the orbital phases of the
asteroid and Jupiter.
Kozai found a remarkable result:  if the asteroid's orbit is sufficiently inclined
(between $39^\circ$ and $141^\circ$), then it cannot remain
on a circular orbit. 
 Instead, its eccentricity and inclination oscillate
periodically.
Furthermore, if the asteroid is 
 highly inclined ($\sim 90^\circ$), 
then its eccentricity will grow over the course
of a single Kozai oscillation to $\sim 1$.

The Kozai mechanism has 
been applied to a variety of astronomical systems.  
For example,  in triple star systems, if the outer star
is inclined it can force Kozai oscillations in the inner 
binary, increasing the inner binary's eccentricity
until tidal dissipation circularizes it
into a very tight orbit \citep[e.g.,][]{EKE01,Dan}.
Similarly, if a Jupiter-mass planet forms around a star
with an inclined companion star, that companion can force
Kozai oscillations in the planet until tides damp its orbit, forming
it into a hot Jupiter \citep{wumurray,Dan}.
Analogous scenarios have been proposed 
for merging supermassive black holes 
\citep{Bla+02} 
and stellar-mass compact objects \citep{Tho10}.
See also \cite{Naoz11} for a review of other applications
of the Kozai mechanism.

In the case considered by Kozai, the asteroid's 
vertical angular momentum (=const$\times \sqrt{1-e^2}\cos i$)
is
 conserved because Jupiter's orbit is circular.
Since this is a two degree of freedom system 
 (eccentricity and inclination) with 
two conserved quantities (vertical angular momentum and secular energy),
it is integrable.
But when Jupiter's orbit is eccentric, the dynamics can no longer be
solved analytically.
As Kozai noted, 
``[w]ithout the aid of a high-speed computer, it is 
rather difficult to estimate the effects of Jupiter's eccentricity.''

The effect 
of an eccentric perturber on Kozai oscillations has been
 considered in a number of  example cases.
 \cite{Har69} integrates the secular octupole equations
 for four triple star systems, and shows that in 
 one case there is a new resonance; in a second case
 there is chaos; and in the other two 
  cases
 no new interesting effects arise.
\cite{holmanetal97} perform  a direct numerical integration for a planet forced
by a star, 
 but with their parameters
 ($\epsilon\simeq 0.008$ in our notation, see below), the
  perturber's finite
 eccentricity
  only leads to a narrow zone of chaos
 around the Kozai separatrix.
\cite{Ford00} integrate  the secular octupole equations
  for a system of two eccentric planets, where the outer planet
  has eccentricity $=0.9$, and find 
 that
 the inner planet's Kozai oscillations are modulated
on long timescales, leading to extreme eccentricities. 
\cite{Bla+02} examine some examples for triples of supermassive
black holes.  \cite{Naoz10,Naoz11} examine some examples
for two planets.  Yet a systematic exploration of the effect of an eccentric
Kozai perturber is lacking.

In a number of studies the
 effect of an eccentric perturber
on Kozai oscillations has been neglected, even in situations
where it should not be \citep{Naoz11}.
Part of the reason is that treatments of Kozai oscillations
often expand the secular Hamiltonian to leading order in the
ratio of semimajor axes, i.e. to quadrupole order.
At that order, the exterior body's argument of periapse does not appear, 
which implies that the particle's
 vertical angular momentum is conserved.\footnote{
The fact that the exterior body's argument of periapse does not appear
to quadrupole order has been called  `` a happy coincidence''
because it makes the system integrable
\citep{lidovziglin76,Las+10}.  However, 
it is perhaps more of an unhappy coincidence 
in view of the fact
 that it has
misled some researchers into neglecting the role of the planet's eccentricity.
}
To see the effect of the eccentricity,
one must work to higher order in the ratio of semimajor axes, i.e. 
at least to octupole order.

In this paper, we 
extend Kozai's work to the case of an eccentric planet and
map out the resulting behavior, which we call the eccentric
Kozai mechanism (EKM).
To do so, we use the
secular octupole Hamiltonian of \cite{Ford00}, which was
apparently first derived by \cite{Harr68thesis} (as cited by
\citeauthor{Har69} \citeyear{Har69}; see also 
 \citeauthor{marchal90} \citeyear{marchal90}
  and 
  \citeauthor{KM99} \citeyear{KM99}
   for derivations).

After the work described  here was completed, we learnt of a similar
paper being prepared by Katz, Dong, and Malhotra.

This paper is organized as follows.
In Section \ref{sec:eom}, 
we present the equations of motion, 
relegating their derivation to the Appendix.  
In Section \ref{sec:quad},
we review the SKM.
In Section \ref{sec:ekm}, the heart of this paper,
we map out the EKM.
We  summarize in Section \ref{sec:conc}.

\section{Equations of Motion}
\label{sec:eom}

We solve for the orbit of a massless test particle in the presence 
of an exterior massive planet,
including only secular interactions expanded to 
octupole order.
The planet is on a fixed eccentric orbit, and
the particle's 
  orbit is specified by four variables,
  \be
\{e,\omega,\theta\equiv \cos i,\Omega\} \ ,
\ee
 which are
its  eccentricity,  argument
of periapse,  inclination (or its cosine), and 
longitude of ascending node relative to the
planet's  periapse \citep[e.g.,][]{MD00}.
In the Appendix, we use the secular octupole
Hamiltonian that has been published in the literature
to derive the particle's equations of motion.
As we show in the Appendix, although that published
Hamiltonian has had its nodes eliminated, one can still use
it to derive the full test particle equations of motion---even 
the equation that requires the nodes.

We summarize the equations here.
Defining the particle's (scaled) total angular momentum and 
vertical angular momentum as%
\footnote{Our variables
are related to the Delaunay variables
 $\{G,g,H,h\}$ 
that are often 
used in treatments of Kozai oscillations via
$J=G/(m\sqrt{G_NM_*a})$, $\omega=g$, $J_z=H/(m\sqrt{G_NM_*a})$, 
$\Omega=h$ (Eqs. [\ref{eq:gbar}]-[\ref{eq:jz2}]). \label{foot}}
\beqn
J\equiv \sqrt{1-e^2}  \\
J_z\equiv \theta\sqrt{1-e^2}  \ ,
\label{eq:jz}
\eeqn
the equations of motion 
may be expressed as partial derivatives
of an energy function $F(e,\omega,\theta,\Omega)$ via
\beqn
{dJ\over dt} &=& {\partial F\over\partial \omega} \label{eq:eom1} \\
{dJ_z\over dt}&=&{\partial F\over\partial \Omega} \\
{d\omega\over dt} &=&{\partial F\over\partial e}{J\over e}+
{\partial F\over\partial\theta}{\theta\over J} \\
{d\Omega\over dt}&=&-{\partial F\over\partial \theta}{1\over J}  \label{eq:eom4}
\eeqn
where $t$ is proportional to time (Eq. [\ref{eq:taudef}]).
These are Hamilton's equations for the two pairs
of canonically conjugate variables  $\{J,\omega\}$ and $\{J_z,\Omega\}$, 
except that 
we  express $F$ as a function of the non-canonical
variables $e$ and $\theta$---for that reason, 
we call $F$ the energy function rather than the Hamiltonian.
Of course, $F$ is a constant of the motion.
It is given by
a quadrupole and an octupole term
\be
F\equiv F_{\rm qu}+\epsilon F_{\rm oc} \ ,
\ee
where the constant
\be
\epsilon\equiv {(a/a_{\rm pl}) e_{\rm pl}\over 1-e_{\rm pl}^2} \  ;
\label{eq:epsdef}
\ee
here $a/a_{\rm pl}$
is the ratio of semimajor axes (inner to outer) and $e_{\rm pl}$
is the planet's eccentricity.
The quadrupole piece is
\be
F_{\rm qu}\equiv
-{e^2\over 2}+\theta^2+{3\over 2}e^2\theta^2+{5\over 2}e^2(1-\theta^2)\cos(2\omega) \ ,
\label{eq:fqu}
\ee
after dropping an irrelevant constant, 
and the octupole term is
\beqn
F_{\rm oc}\equiv {5\over 16}(e+{3\over 4}e^3)
\Big[
	\left(
		1+11\theta-5\theta^2-15\theta^3
	\right)\cos(\omega+\Omega)  \nonumber \\
  +    \left(
              1-11\theta-5\theta^2+15\theta^3
       \right)\cos(\omega-\Omega) \nonumber
\Big] \nonumber \\
-{175\over 64}e^3
\Big[
	\left(
		1-\theta-\theta^2+\theta^3
	\right)    \cos(3\omega-\Omega) \ \ \ \  \nonumber \\
+	\left(
		1+\theta-\theta^2-\theta^3
	\right) \cos(3\omega+\Omega)
\Big] \ \ \ \ \ .
\label{eq:foct}
\eeqn

The only adjustable parameter in the equations of motion other than 
the initial conditions  is 
the constant
$\epsilon$.  That constant encodes the properties of the planet.\footnote{
The planet's properties also enter in a trivial way through the scaling between $t$
and time (Eq. [\ref{eq:taudef}]).
}
In the standard Kozai mechanism (SKM), the planet's orbit is circular
($\epsilon=0$).
Hence
$F$ is independent of $\Omega$, and thus 
$J_z$ is a constant of the motion.
But if the planet's eccentricity is not zero ($\epsilon>0$), then the
 octupole term allows $J_z$ to change.
If the planet is either nearly circular ($e_{\rm pl}\ll 1$) or distant
($a/a_{\rm pl}\ll1 $), then $\epsilon$ is very small, and the
evolution is typically similar to the SKM (except at extremely high
inclinations).  But as we shall show, if $\epsilon \gtrsim 0.01$, then
the octupole term
 can lead to qualitatively new behavior
that is not found in the $\epsilon=0$ limit.

\section{The Standard Kozai Mechanism ($\epsilon=0$)}
\label{sec:quad}

\begin{figure}
\centerline{\includegraphics[width=0.5\textwidth]{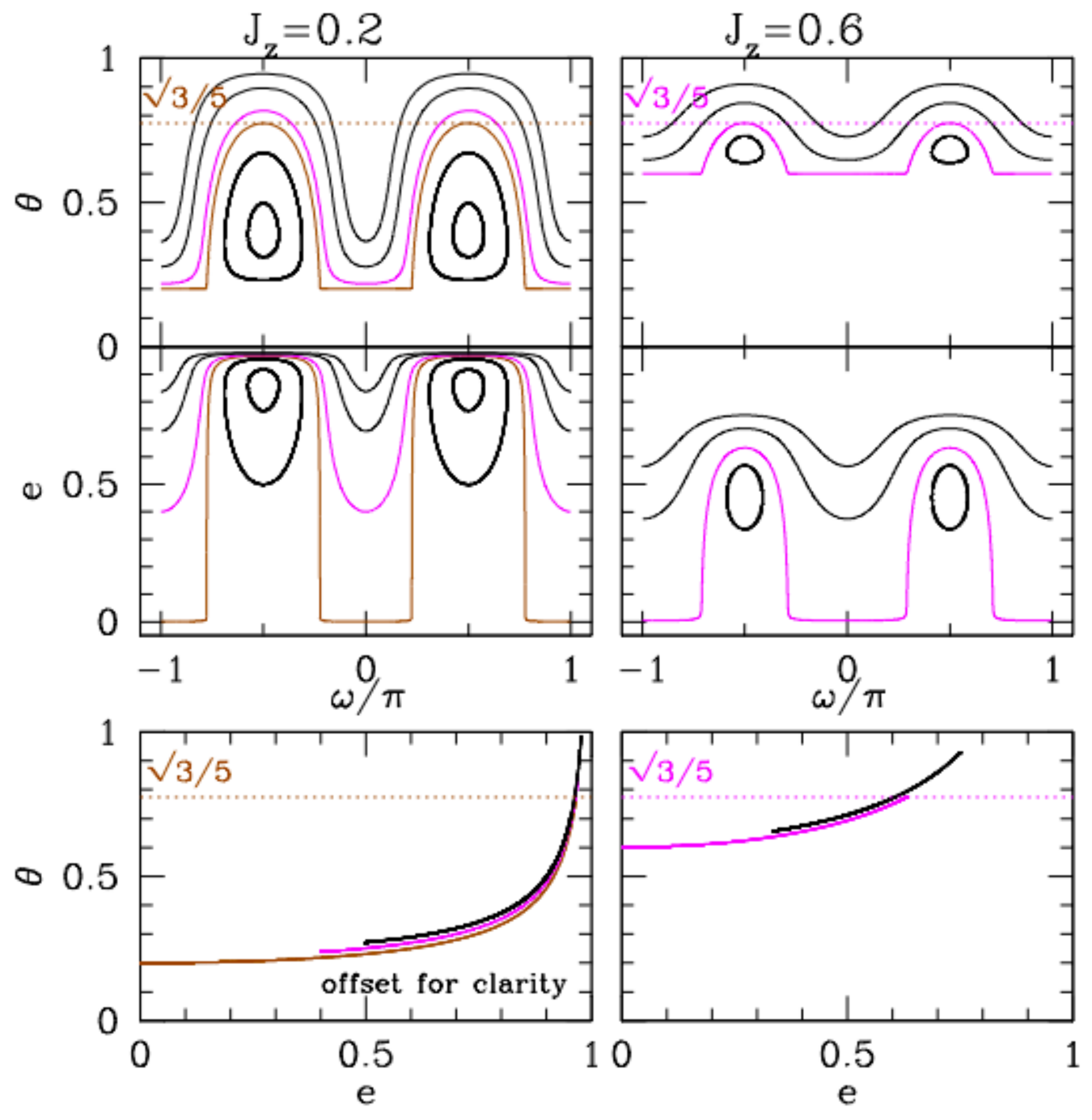}}
\caption{SKM ($\epsilon=0$): The 
three left panels show trajectories with $J_z=0.2$ and various
values of $F$.  Trajectories with the same $J_z$
 overlap
in the $e$-$\theta$ plane (bottom panel); we have slightly offset these for
clarity.
 The right panels show the same, but with $J_z=0.6$.
 For circulating trajectories, the minimum $e$ and $\theta$
occur at $\omega=0$.
 The dashed horizontal lines show the cosine of the critical
 Kozai inclination, $\sqrt{3/5}$.  The separatrix---which
  separates circulating from librating trajectories---always
has
 $e_0=0$ and $|\theta_{\rm \pi/2}|=\sqrt{3/5}$}
\label{fig:quad} 
\end{figure}

We review the SKM 
 to set the stage for the EKM.
For the SKM, the planet's orbit is circular ($\epsilon=0$, i.e.
$F=F_{\rm qu}$).  Hence
 there are two constants of the motion, 
$F$ and $J_z$, and the motion is regular. 
Each trajectory
may be labelled by the values of $F$ and $J_z$. 
Inserting Equation (\ref{eq:jz}) into Equation (\ref{eq:fqu})
immediately determines $e$ as a function of $\omega$, as
well as $\theta$ as a function of $\omega$.
The three left panels of Figure \ref{fig:quad} show sample trajectories,
with fixed $J_z=0.2$ and various values of $F$; the right panels
show the same, but with $J_z=0.4$.
For our purposes, the most important properties of the SKM are as follows:
\bi
\item
	Because $J_z=$constant, each trajectory traces
	out a curve in the $e$-$\theta$ plane, which we call a 
	``Kozai curve.''
\item
	There are two classes of trajectories, librating and circulating.
	On  circulating trajectories, $e$ and $|\theta|$ are smallest  
	at $\omega=0$, and they are largest
	at $\omega=\pm \pi/2$.  
	The  separatrix  has 
	$e_0\equiv e\vert_{\omega=0}=0$\footnote{
	Throught this paper, we denote values at $\omega=0$
	with subscript '0'.
	} and $|\theta_{\pm \pi/2}|=\sqrt{3/5}$.
\item	
	On a trajectory that has $e_0\ll 1$, the largest
	$e$ is 
	\be
	 e^2_{\pm \pi/2}\approx 1-{5\over 3}\theta_0^2
	\approx 1-{5\over 3}J_z^2 \ ,
	\label{eq:emax}
	\ee
	to leading order in $e_0^2$, when $|\theta_0|<\sqrt{3/5}$.
	(The corresponding  inclinations are the critical Kozai angles
	$\cos^{-1}\pm\sqrt{3/5}=39^\circ$ and  $141^\circ$.)
	Therefore when $|\theta_0|\ll 1$ (i.e. inclination close to $90^\circ$), 
	the largest eccentricity is nearly unity.
\item
	Given $F$ and $J_z$, the minimal $e$ on a circulating trajectory
	satisfies
	\be
	e_0^2={1\over 2}(F-J_z^2) \ . \label{eq:e02} 
	\ee
\ei

\begin{figure}
\centerline{\includegraphics[width=0.5\textwidth]{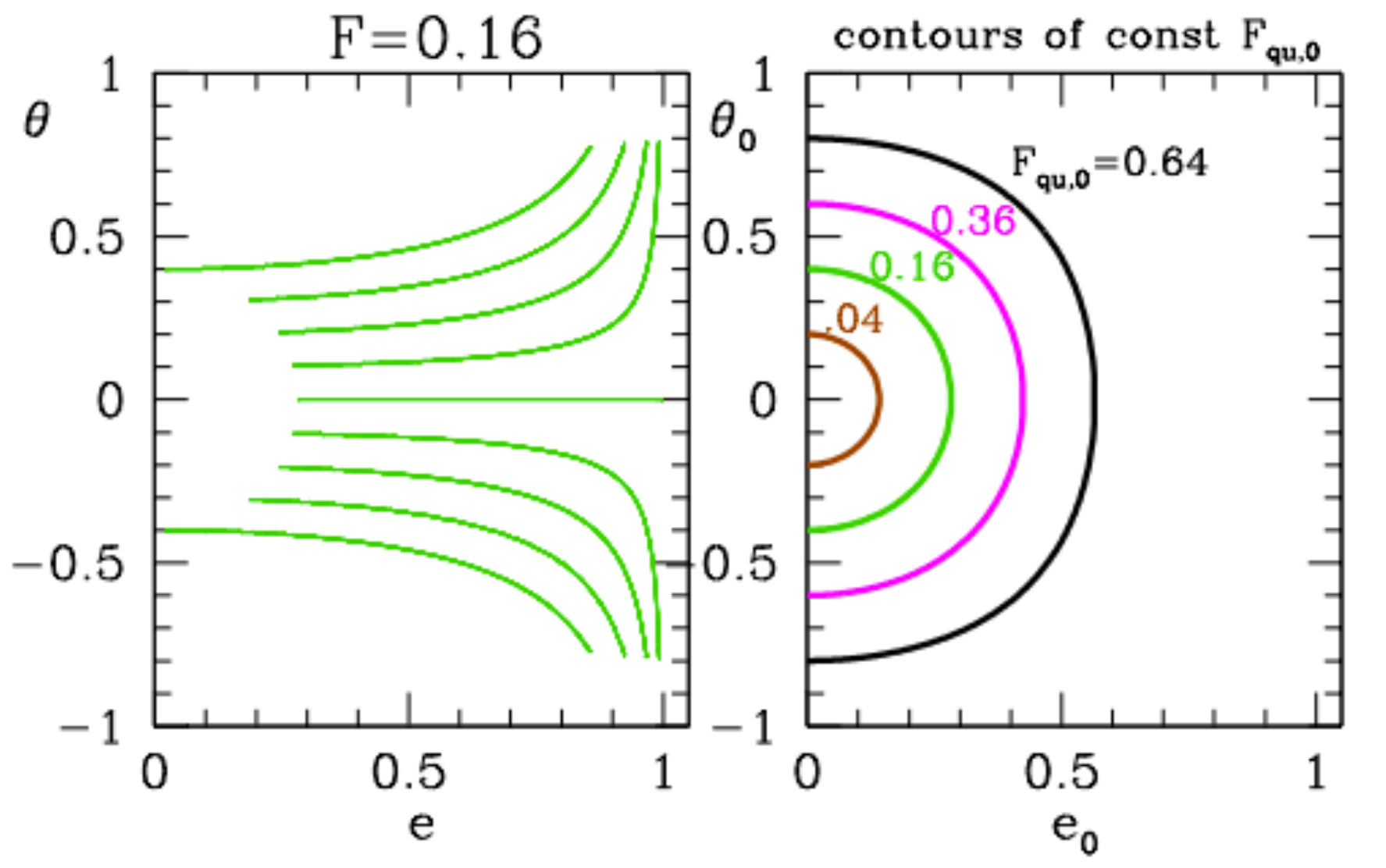}}
\caption{SKM ($\epsilon=0$): 
The left panel shows trajectories with fixed $F$ and various
values of $J_z$ (``Kozai curves'').
The left-most tips of the Kozai curves
(with co-ordinates $\{e_0,\theta_0\}$)
 trace out another curve, 
which we call the ``energy curve.''
That energy curve is shown in green in the right panel; it
is the level curve of  $F_{\rm qu,0}\equiv F_{\rm qu}\vert_{\omega=0}=0.16$.
The right panel also shows some other energy curves.  These intersect the 
$\theta_0$ axis at $\sqrt{F_{\rm qu,0}}$ and the $e_0$ axis at
 $\sqrt{F_{\rm qu,0}/2}$
\label{fig:quad2}}
\end{figure}

Figure \ref{fig:quad2}, left panel,  shows a sequence of 
Kozai curves with the same
$F$ and differing $J_z$. 
 The left boundaries of the sequence
 (i.e., the values $\{e_0,\theta_0\}$ for each Kozai curve)
  trace out
a curve in the $e$-$\theta$ plane, 
which we call an ``energy curve.''
That curve is plotted in green in the right panel
of Figure \ref{fig:quad2}.  
An energy curve
is a curve of constant  $F_{\rm qu}\vert_{\omega=0}$.
Its form is given by Equation
(\ref{eq:e02}) with $J_z^2=e_0^2(1-\theta_0^2)$. 
The right panel of Figure \ref{fig:quad2} also 
shows other energy curves with different energies
$F_{\rm qu}$.

\section{The Eccentric Kozai Mechanism ($\epsilon> 0$)}
\label{sec:ekm}

\begin{figure}
\centerline{\includegraphics[width=0.5\textwidth]{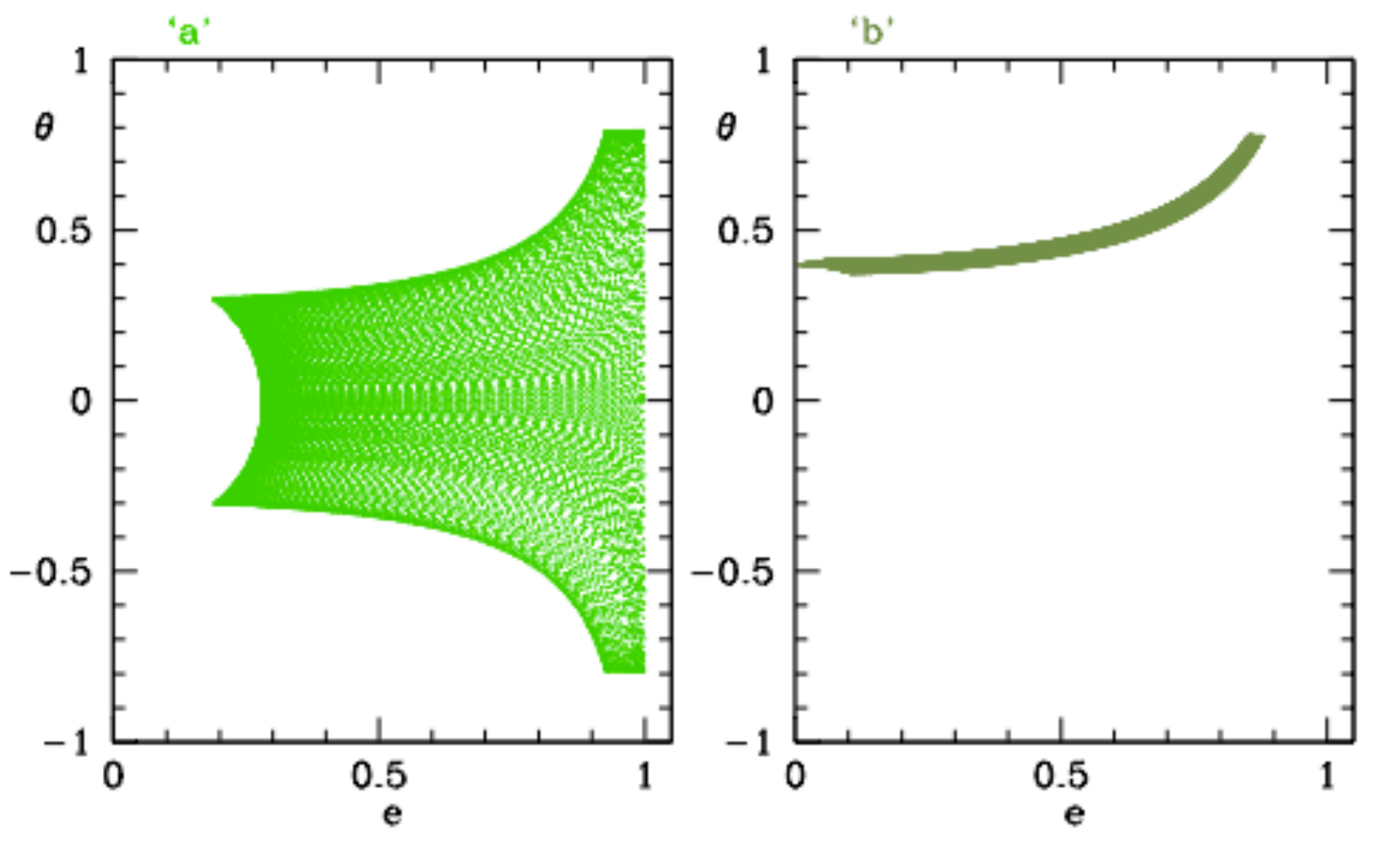}}
\caption{EKM: 
Two sample trajectories (`a' and `b') with  $F=0.16$, $\epsilon=0.01$, and different
initial conditions.
When the planet is eccentric, the test particle no longer traces out a single
Kozai curve, but evolves from one Kozai curve to another, 
approximately tracing out a sequence of
Kozai curves that have a fixed quadrupole energy (Fig. \ref{fig:quad2}, left panel).
Orbit `a' flips repeatedly (i.e. crosses $\theta=0$) and reaches extreme eccentricities,
 and orbit `b' does not. 
Although only points up to time $t=500$ are plotted, longer runs
yield identical plots, albeit with more densely packed points.
In particular,  orbit `b' never flips.
\label{fig:traj1}
}
\end{figure}

When the planet is eccentric  ($\epsilon> 0$) there
 is only a single conserved quantity, the 
secular energy $F$. 
  Therefore the particle's trajectories 
are more complicated, and can even be chaotic.
Figure \ref{fig:traj1} shows two sample trajectories in the 
$e$-$\theta$ plane for $\epsilon=0.01$.  Both trajectories
have the same energy $F=0.16$, but different initial conditions
consistent with that energy.
Rather than being confined to a single 
Kozai curve, the particle evolves from one Kozai curve
to another.  
As long as $\epsilon\ll 1$, all of these Kozai curves
have nearly the same $F_{\rm qu}$, and therefore
the trajectories in Figure \ref{fig:traj1} follow along the tracks
displayed in the left panel of Figure \ref{fig:quad2}.

Trajectory `a' in Figure \ref{fig:traj1}  evolves through $\theta=0$.  
Its orbit flips from prograde
($\theta>0$) to 
retrograde ($\theta<0$) and back again.
Furthermore, its eccentricity approaches unity.
In fact, the flipping of an orbit is closely tied to its
eccentricity reaching unity.
This can be seen from Equation (\ref{eq:emax}), 
which implies that a Kozai curve that has $\theta_0\approx 0$
reaches $e\approx 1$. (It never reaches exactly $e=1$; see below).
Trajectory `b' never flips, and its eccentricity does not approach
unity.

\begin{figure}
\centerline{\includegraphics[width=0.5\textwidth]{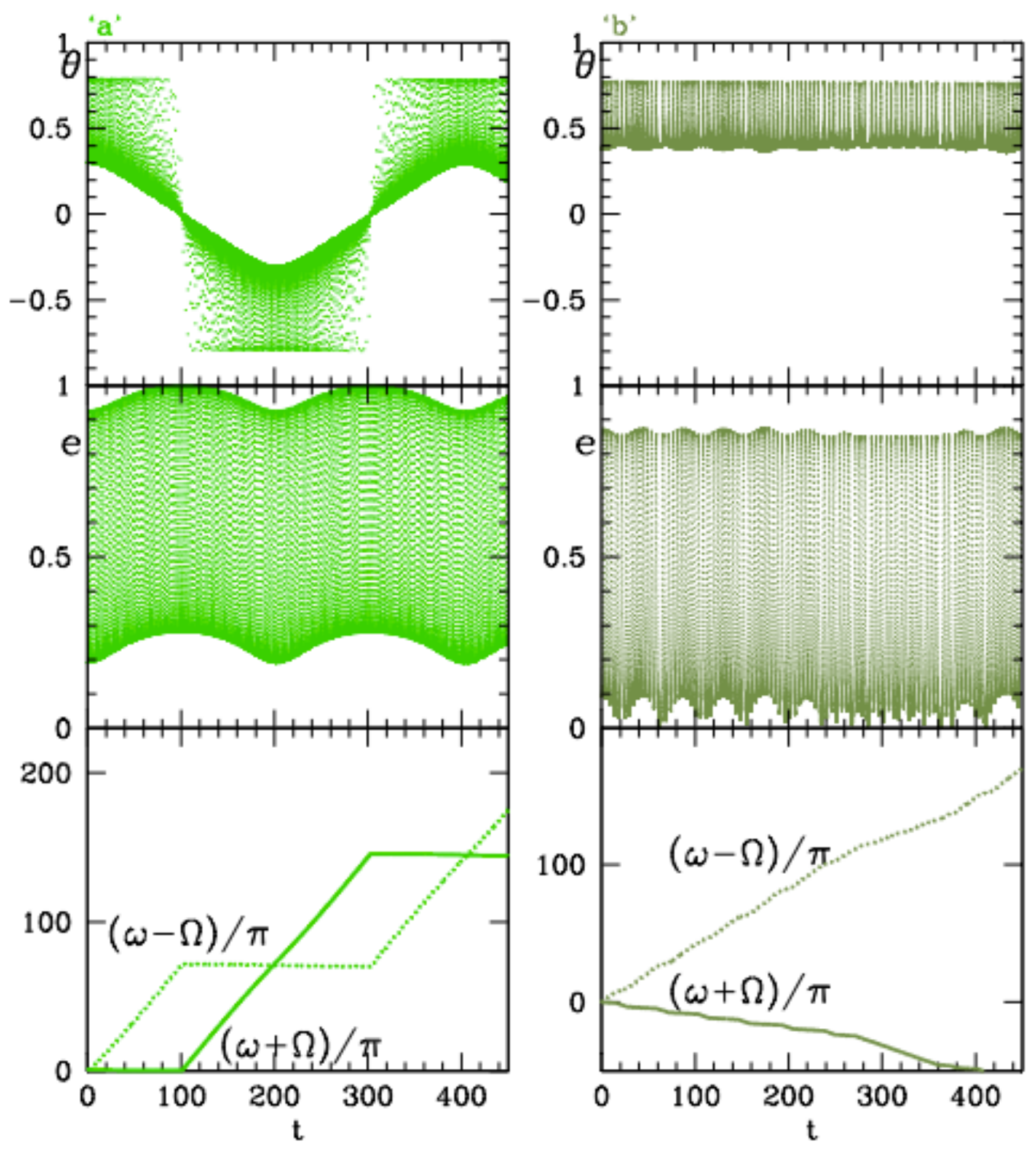}}
\caption{EKM: 
Temporal evolution of the two trajectories depicted in Figure \ref{fig:traj1}.
In the bottom panels, the angles are cumulative.
Trajectory `a' reaches extreme eccentricities when it flips.
Its evolution is regular, with $\omega+\Omega$ librating 
when it is prograde, and $\omega-\Omega$ librating when it is
retrograde.  Trajectory `b' is chaotic, and never flips.
\label{fig:traj2}
}
\end{figure}

The left panels of Figure \ref{fig:traj2} show the temporal evolution 
of $\theta$ and $e$ for trajectory `a'.  The bottom left panel 
shows  two angles 
that appear as arguments of cosine terms
in $F_{\rm oc}$.  When the particle's  orbit is prograde, 
the angle $\omega+\Omega$ librates and $\omega-\Omega$ circulates;
and when it is retrograde, those two angles switch roles.
The right panels of Figure \ref{fig:traj2} show the same quantities
for trajectory `b'.  
Whereas trajectory `a' is regular, `b' is chaotic.  We 
demonstrate that more explicitly below.

\begin{figure}
\centerline{\includegraphics[width=0.5\textwidth]{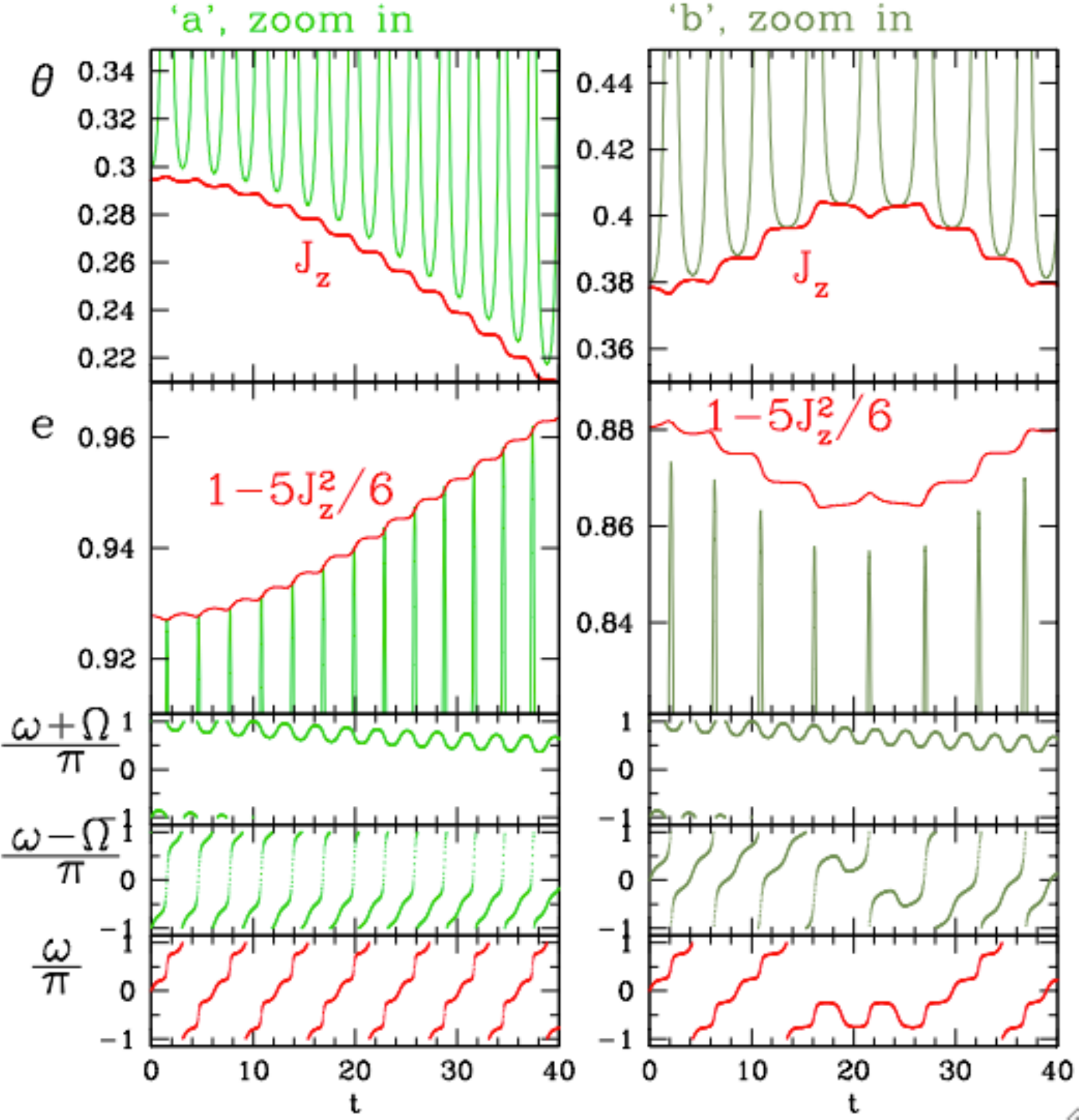}}
\caption{EKM: 
 Zoomed in evolution of trajectories `a' and `b' at early times (Fig. \ref{fig:traj1}).
In the top panels the red lines show that $J_z$ controls
 the envelope of both $\theta$ and $e$ (via eq. [\ref{eq:emax}]).
In the bottom panels, the angles are not cumulative.
 \label{fig:traj3}
 }
\end{figure}

The left panels of Figure \ref{fig:traj3} are a zoom-in of the `a' trajectory 
at early times.
 The fast oscillations in $\theta$ and $e$ are primarily governed
 by the quadrupole evolution, 
as $e$ and $\theta$
trace out Kozai  curves with $F=0.16$ and various values of $J_z$.
Over the course of a single oscillation, 
$\omega$ increases from $0$ to $\pi$ or from $-\pi$ to 0 . Hence the orbit is always circulating (see
Fig. \ref{fig:quad}).
The top left panel also shows $J_z$.  
Whereas in the SKM
$J_z=$const., here $J_z$ changes in a nearly step-wise fashion.  
There are sharp jumps in $J_z$ whenever $e$ and $\theta$
change rapidly; these are forced by the octupolar contribution $F_{\rm oc}$.
The long-term evolution of $J_z$ controls the  envelopes of both
 $\theta$ (Fig. \ref{fig:traj3}, top-left panel) and $e$ (Fig. \ref{fig:traj3}, middle-left
 panel; see eq. [\ref{eq:emax}]). 
Successive maxima of $e$ occur in discrete steps.  
Therefore even when $J_z$ crosses through zero, the maximum
$e$ is never precisely unity.  Nonetheless, as time evolves,
 the maximum $e$ reached  approaches closer and closer to unity.

The right panels of Figure \ref{fig:traj3} show the corresponding
zoom-in for trajectory `b'.  
 The bottom right panel shows that  $\omega$ switches
from circulation to libration and back again.  That explains why
 the `b' trajectory is chaotic.  Similar behavior is shown in 
 \cite{holmanetal97}.

\begin{figure}
\centerline{\includegraphics[width=0.5\textwidth]{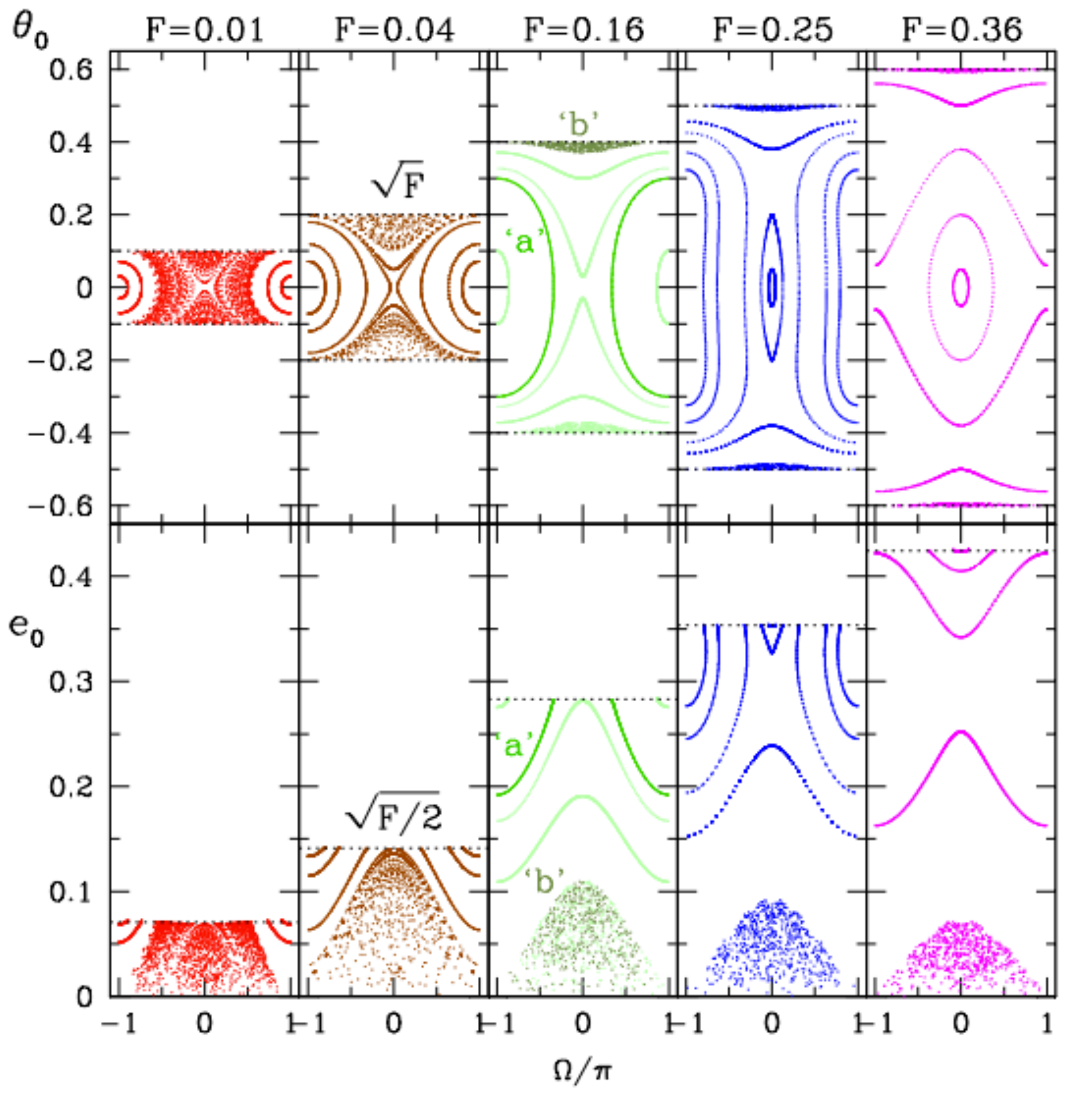}}
\caption{EKM, $\epsilon =0.01$.  Surfaces of section
with various values of $F$; points are plotted whenever
 $\omega=2n\pi $ for integer $n$. 
 For each $F$, if the values of $\theta_0$ were plotted against
 the value of $e_0$, they would lie along an energy curve (Fig. \ref{fig:quad2}, 
 right panel). The colors in that figure correspond to the colors in this one.
\label{fig:sos}
}
\end{figure}

A global view of the dynamics is provided by surfaces of section.
Figure \ref{fig:sos} maps out the behavior when
$\epsilon=0.01$.  
We plot a point whenever $\omega=2n\pi$
for integer $n$.\footnote{Trajectories
only appear on our sections while $\omega$ is circulating, because
 $\omega$ is never equal to $2n\pi$ during librations (Fig. \ref{fig:quad}).}
This may be interpreted as follows.  Consider trajectory `a' of Figure \ref{fig:traj1}.
When $\omega=2n\pi$ that trajectory hits  the green energy curve depicted
in the right panel of  Figure \ref{fig:quad2}.
At those times we plot in Figure \ref{fig:sos} the
value of $e=e_0$ versus $\Omega$
and $\theta=\theta_0$ versus $\Omega$
(curves labelled `a', middle panels). 
Also shown in those green middle panels 
are other trajectories with the same energy $F=0.16$, 
including trajectory `b'.
If  all of the green points were plotted against each other
in the $e_0$-$\theta_0$ plane, they would trace out the energy
curve labelled 0.16 in Figure \ref{fig:quad2}. 
Equivalently, one can imagine that that energy curve is extended
out of the plane of the paper into a half-cyliner, with the third
 dimension being the value
of $\Omega$.  The $F=0.16$ surfaces of section of Figure \ref{fig:sos}
cover the surface of that half-cylinder.  Of course, for a given 
energy, the maximum $|\theta_0|$ is $\sqrt{F}$, and the maximum 
$e_0$ is $\sqrt{F/2}$.  Beyond those values, the energy curve does
not exist (Fig. \ref{fig:quad2} caption).

The other panels in Figure \ref{fig:sos} may be interpreted similarly, 
with each pair of panels at  fixed $F$ lying along a single energy
curve of Figure \ref{fig:quad2} (with corresponding colors).  
We therefore now have a virtually complete view of the $\epsilon=0.01$
dynamics.

Perhaps the most dramatic new behavior caused by the  octupole 
term  is 
that orbits can flip
 orientation (i.e. cross
$\theta=0$), and, as a consequence, reach arbitrarily 
high values of eccentricity.
From Figure \ref{fig:sos}, it is immediately clear which 
orbits exhibit this behavior, and which do not, at $\epsilon=0.01$.  
For example, orbits with $e_0=0$ will always flip provided that
$|\theta_0|\lesssim 0.2$  (i.e., $i_0\gtrsim \cos^{-1}0.2\sim 80^\circ$ for
prograde orbits).  
For $e_0\ne 0$, we infer from the blue ($F=0.25$) panels
that even orbits with $\theta_0$  as large as 0.4 
($i_0$ as small as $66^\circ$) can flip.

A second new behavior is the appearance of chaos.
Regular orbits appear as curves in the surfaces of section, while
chaotic orbits appear as a smattering of points.  
For example,  it is apparent 
from Figure \ref{fig:sos} that trajectory `a' is regular and `b' is chaotic.
Chaos always occurs near $e_0=0$.  That is because
 the Kozai separatrix always has $e_0=0$ 
(Section \ref{sec:quad}), and 
 chaos is caused the crossing of the Kozai separatrix, 
when $\omega$ transitions from librating to circulating and 
back to librating (e.g., Fig. \ref{fig:traj3}).

\begin{figure}
\centerline{\includegraphics[width=0.5\textwidth]{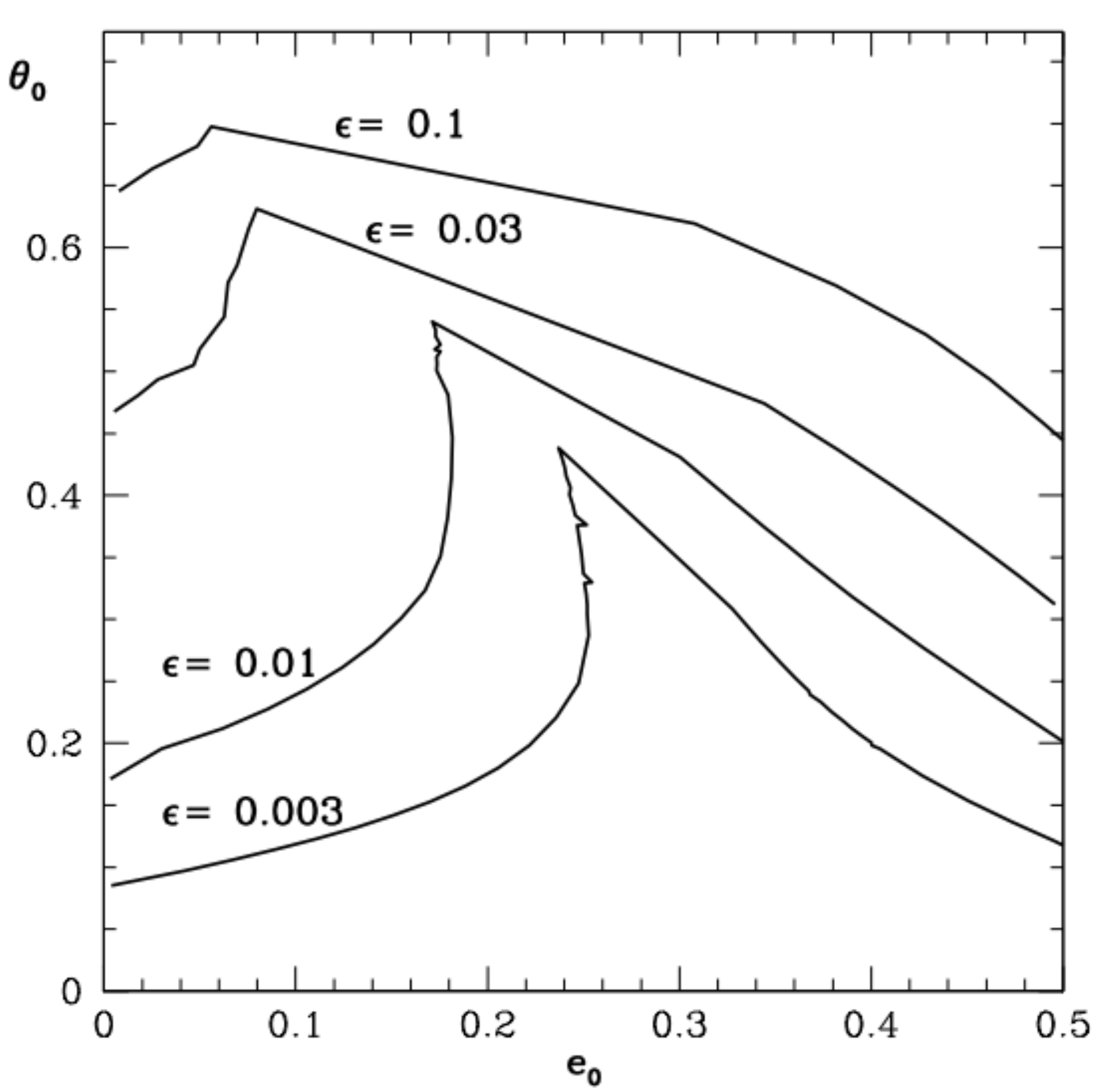}}
\caption{First Flipping Orbit.  
Each curve marks the first flipping orbit
that can occur for a given value of $\epsilon$.
Above each curve (i.e., at low inclination)
none of the orbits flip, and below it some do.
}
\label{fig:ffo}
\end{figure}

Figure \ref{fig:ffo} is the main result of this paper. 
It summarizes where flipping orbits occur for various
values of $\epsilon$. 
Towards the top of the plot (large $\theta_0$ and
small $i_0$), there are no flipping orbits.  
But as $\theta_0$ decreases, an increasing number of orbits
can flip.  Each curve marks, for a given $\epsilon$,
the location at which the first flipping orbit occurs, if
$\Omega$ is judiciously chosen.  Above the
curve there are no flipping orbits, and below there are some---typically
around half of the orbits below the curve can flip.
We made this plot by running a suite of numerical simulations initialized
with $\theta=0$ and $\omega=0$, and scanning through values of initial $e$
and $\Omega$ to determine which flipping orbit reached the largest values
of $\theta_0$ for a given initial $e$, and then plotting the extreme values 
of $e_0$ and $\theta_0$ for that
orbit.   (Note that all such trajectories initialized with the same initial $e=e_0$ share the same energy curve.)

A striking feature of Figure \ref{fig:ffo} is the fairly sudden transition 
when $\epsilon$ exceeds  a few percent.  
For example, focusing on orbits with $e_0=0$, when $\epsilon =0.01$
flipping orbits can occur only if $|\theta_0|\lesssim 0.2$
 i.e., $i_0\gtrsim 80^\circ$ (for prograde), as stated above.
 But for $\epsilon=0.03$ flips can occur for $|\theta_0|\sim 0.5$, 
 i.e. $i_0\sim 60^\circ$ (for prograde).  Note that if test particle orbits are randomly 
 oriented, they will be uniformly distributed in $\theta$, and therefore
 half of the orbits will have $\theta<0.5$.  Of course, this is only a rough 
 estimate for the fraction of flipping orbits, since it depends also on the values
 of the other variables ($e,\omega,\Omega$).
 
The curves in Figure \ref{fig:ffo} rise as $e_0$ increases from 0, and then 
suddenly turn over.  The reason for the sudden turn over
 is apparent from a glance at the
surfaces of section (Fig. \ref{fig:sos}).  In the blue ($F=0.25$) section, there
are flipping orbits centered both on $\Omega=0$ and on $\Omega=\pi$, with
the latter ones reaching to a larger $|\theta_0|$.  But in the magenta ($F=0.36$) section, 
there are no more flipping orbits centered at $\Omega=\pi$.  It is that
discontinuity that gives rise to the sharp break in Figure \ref{fig:ffo}.

\section{Summary}
\label{sec:conc}

In this paper, we map out the behavior of a test particle that is forced
by a massive eccentric planet,  working under the secular octupole approximation.

In Section \ref{sec:eom}, we give the evolutionary equations.
These are derived in the Appendix, where we show
how one may use the published Hamiltonian---which has had its nodes
eliminated---to derive the full equations of motion.
There is a single parameter, $\epsilon$, that characterizes
deviations from the SKM
 (Eq. [\ref{eq:epsdef}]).  
 The SKM is recovered 
 when $\epsilon\ll 1$, i.e. when the planet
 is either nearly circular or very distant.

In Section \ref{sec:quad}, we review the SKM.
In the $e$-$\theta$ plane (where $\theta\equiv \cos i$), the particle
traces out a Kozai curve because its $J_z$ is fixed.  
A sequence of Kozai curves with the same energy $F$ outline an 
energy curve, which is a curve of constant secular energy, evaluated
at $\omega=0$.

In Section \ref{sec:ekm}, we map out what happens
when the planet's orbit is eccentric.  
Our principal results are:
\bi
\item
In the $e$-$\theta$ plane, a single trajectory
evolves from one Kozai curve to another,
 all of which abut the same energy curve (Fig. \ref{fig:traj1}).
\item
For trajectories where the orbit flips, the eccentricity becomes extremely high, and
as time evolves, the maximum eccentricity reached approaches closer
and closer to unity. 
That is because every time $\theta$ crosses through zero,  $|\theta_0|$ becomes
very small. And given enough zero-crossings,  the smallest $|\theta_0|$
reached becomes arbitrarily small, and hence the maximum eccentricity 
becomes arbitrarily close to unity (as determined by  Eq. [\ref{eq:emax}]).
\item
The vertical angular momentum $J_z$ varies gradually in the EKM, unlike in the 
SKM where it is constant.
Its evolution controls that of both $\theta$ and $e$ (Fig. \ref{fig:traj2}).
Even when the value of $\epsilon$ is small, and hence the temporal evolution
of $J_z$
is slow, if $J_z$ crosses through zero, the dramatic flipping and extreme eccentricity
behavior occurs (Fig. \ref{fig:traj1}, left panel).
\item
Chaotic behavior occurs when the Kozai separatrix is crossed.  In that
case, $\omega$ transitions from librating to circulating, and back.  See also
 \cite{holmanetal97}.

\item
Surfaces of section, taken whenever $\omega=2n\pi$, provide
an graphical global map of the dynamics.
Each energy curve gives rise to 
 two panels in Figure \ref{fig:sos} (top and bottom).  By stepping through the energy curves, one acquires a complete map 
 of the dynamics for a given value of $\epsilon$. 
From the surfaces of section, one can immediately see which orbits
exhibit flipping and extreme eccentricity, and which do not.  
One can also see which orbits are chaotic and which are not.  All four combinations
 occur: flipping and non-chaotic (`a'), flipping and chaotic, non-flipping 
and non-chaotic, non-flipping and chaotic (`b').

\item
Figure \ref{fig:ffo} is the main result of this paper.  For each 
$\epsilon$, it gives the curve in the $e_0$-$\theta_0$ plane at 
which the first flipping orbit occurs (along with the accompanying
extreme-eccentricity behavior).  Above the curve no flipping orbits occur, 
and below it an increasing
number occur.  
There is a fairly sharp transition in the smallest inclination at which orbits
can flip.  For $\epsilon\sim 0.01$,  the inclination must 
exceed $\sim 80^\circ$ to flip (for prograde orbits with $e_0=0$);
 but when $\epsilon\sim 0.03$, 
it need only exceed $\sim 60^\circ$, implying that around half of randomly 
oriented test particle orbis will flip in that case.

\ei

We have made a number of approximations
in order to simplify the analysis. 
First, we work under the secular octupole approximation, 
which is poor when $\epsilon$
is too large because in that case the particle comes close to the planet.
We have made  a few comparisons with direct integrations, 
some of which are presented in \cite{Naoz11}. Thus
far the agreement is quite good, but
 more should be done to determine 
the domain of validity of the secular octupole approximation.
Second, we take the inner orbiting body to be massless.
While that is appropriate for a planet being forced by a companion 
star, it is not for two planets, which is  the main motivation for this paper
\citep{Naoz10}.
We are currently
 extending the work presented
here to two massive planets; in that case, there is a generalized
$\epsilon$ that includes the masses \citep{Naoz11}. 
And third, we ignore all physical effects beyond
Newtonian gravity, including tides and general relativistic
precession.  These will play a role when the inner planet's eccentricity
approaches unity.



\section*{Acknowledgments}
We thank W. Farr,  B. Katz, F. Rasio, J. Teyssandier, and Y. Wu for helpful
discussions.
SN acknowledges support from the Gruber Foundation
  Fellowship and from the National Postdoctoral Award Program for
  Advancing Women in Science (Weizmann Institute of
  Science).
\newline

\bibliographystyle{apj}

\bibliography{tp}

\begin{thebibliography}{19}
\expandafter\ifx\csname natexlab\endcsname\relax\def\natexlab#1{#1}\fi

\bibitem[{{Blaes} {et~al.}(2002){Blaes}, {Lee}, \& {Socrates}}]{Bla+02}
{Blaes}, O., {Lee}, M.~H., \& {Socrates}, A. 2002, \apj, 578, 775

\bibitem[{{Eggleton} \& {Kiseleva-Eggleton}(2001)}]{EKE01}
{Eggleton}, P.~P. \& {Kiseleva-Eggleton}, L. 2001, \apj, 562, 1012

\bibitem[{{Fabrycky} \& {Tremaine}(2007)}]{Dan}
{Fabrycky}, D. \& {Tremaine}, S. 2007, \apj, 669, 1298

\bibitem[{{Ford} {et~al.}(2000){Ford}, {Kozinsky}, \& {Rasio}}]{Ford00}
{Ford}, E.~B., {Kozinsky}, B., \& {Rasio}, F.~A. 2000, \apj, 535, 385

\bibitem[{{Ford} {et~al.}(2004){Ford}, {Kozinsky}, \& {Rasio}}]{Ford00error}
---. 2004, \apj, 605, 966

\bibitem[{{Harrington}(1968)}]{Harr68thesis}
{Harrington}, R.~S. 1968, PhD thesis, THE UNIVERSITY OF TEXAS AT AUSTIN.

\bibitem[{{Harrington}(1969)}]{Har69}
---. 1969, Celestial Mechanics, 1, 200

\bibitem[{{Holman} {et~al.}(1997){Holman}, {Touma}, \&
  {Tremaine}}]{holmanetal97}
{Holman}, M., {Touma}, J., \& {Tremaine}, S. 1997, \nat, 386, 254

\bibitem[{{Kozai}(1962)}]{Kozai}
{Kozai}, Y. 1962, \aj, 67, 591

\bibitem[{{Krymolowski} \& {Mazeh}(1999)}]{KM99}
{Krymolowski}, Y. \& {Mazeh}, T. 1999, \mnras, 304, 720

\bibitem[{{Laskar} \& {Bou{\'e}}(2010)}]{Las+10}
{Laskar}, J. \& {Bou{\'e}}, G. 2010, \aap, 522, A60+

\bibitem[{{Lidov} \& {Ziglin}(1976)}]{lidovziglin76}
{Lidov}, M.~L. \& {Ziglin}, S.~L. 1976, Celestial Mechanics, 13, 471

\bibitem[{{Marchal}(1990)}]{marchal90}
{Marchal}, C. 1990, {The three-body problem}, ed. {Marchal, C.}

\bibitem[{{Murray} \& {Dermott}(2000)}]{MD00}
{Murray}, C.~D. \& {Dermott}, S.~F. 2000, {Solar System Dynamics} (Cambridge
  University Press)

\bibitem[{{Naoz} {et~al.}(2011{\natexlab{a}}){Naoz}, {Farr}, {Lithwick},
  {Rasio}, \& {Teyssandier}}]{Naoz10}
{Naoz}, S., {Farr}, W.~M., {Lithwick}, Y., {Rasio}, F.~A., \& {Teyssandier}, J.
  2011{\natexlab{a}}, \nat, 473, 187

\bibitem[{{Naoz} {et~al.}(2011{\natexlab{b}}){Naoz}, {Farr}, {Lithwick},
  {Rasio}, \& {Teyssandier}}]{Naoz11}
---. 2011{\natexlab{b}}, submitted

\bibitem[{{Thompson}(2010)}]{Tho10}
{Thompson}, T.~A. 2010, ArXiv e-prints

\bibitem[{{Triaud} {et~al.}(2010){Triaud}, {Collier Cameron}, {Queloz},
  {Anderson}, {Gillon}, {Hebb}, {Hellier}, {Loeillet}, {Maxted}, {Mayor},
  {Pepe}, {Pollacco}, {S{\'e}gransan}, {Smalley}, {Udry}, {West}, \&
  {Wheatley}}]{Tri+10}
{Triaud}, A.~H.~M.~J., {Collier Cameron}, A., {Queloz}, D., {Anderson}, D.~R.,
  {Gillon}, M., {Hebb}, L., {Hellier}, C., {Loeillet}, B., {Maxted}, P.~F.~L.,
  {Mayor}, M., {Pepe}, F., {Pollacco}, D., {S{\'e}gransan}, D., {Smalley}, B.,
  {Udry}, S., {West}, R.~G., \& {Wheatley}, P.~J. 2010, \aap, 524, A25+

\bibitem[{{Wu} \& {Murray}(2003)}]{wumurray}
{Wu}, Y. \& {Murray}, N. 2003, \apj, 589, 605

\end{thebibliography}

\appendix
\section{\bf   Octupole Secular Equations of Motion For a Test Particle}

The secular interaction energy between an interior and exterior planet
(unprimed and primed, respectively) can be written as
\be
E_{\rm eon}=-G_N{mm'\over a'}\alpha^2 f(\alpha,e',e,g',g,\theta) \ ,
\label{eq:ee}
\ee
where 
$\theta\equiv \cos I $
and the other quantities are the mutual inclination $I$, 
argument of periapse $g$, eccentricity $e$, 
semimajor axis $a$, ratio of semimajor axes (inner to outer)
$\alpha$, mass $m$, and Newton's constant $G_N$.
(The notation here differs slightly from  the body of 
the paper in order to connect to previous
treatments of the octupole.  See footnote \ref{foot} for the relation
between the two notations.)
The subscript  ``eon'' refers to the elimination
of the nodes, 
i.e. the longitudes of ascending nodes ($h,h'$) have been chosen to satisfy
$h-h'=\pi$
which is why they do not appear in the energy.

We seek the equations of motion for the inner
planet in the limit that  $m\rightarrow 0$.
The equations of motion can be expressed as Hamilton's equations for the canonical 
Delaunay variables
$\{g,G; h,H\}$ where
\beqn
G \equiv m\sqrt{G_NM_* a}\sqrt{1-e^2}  \ , \ \ \ 
H\equiv  G\cos i \ ,
\eeqn
with $M_*$ the mass of the central star.
Note that  in secular theory, $a$ is constant; note also that
in the test particle limit, $i=I$.

Setting the Hamiltonian ${\cal H}=
E_{\rm eon}$ will give the correct equations of motion 
of $g,G,$ and $h$, but not for $H$ because the nodes 
have been eliminated
\citep{Naoz11}.
 We shall prove that the
correct procedure for obtaining the Hamiltonian 
is simply to replace $g'\rightarrow \pi-h$ in the above
expression for $E_{\rm eon}$, as well as
$e'=$const., $e^2\rightarrow 1-G^2/(m^2G_NM_\odot a)$, 
and $\theta\rightarrow H/G$.  
The relation $g'\rightarrow \pi-h$ might appear surprising
because one might have expected that $g'=$const when the
inner particle is massless.  But in truth $h'$ is undefined because
the reference plane is aligned with the outer orbit.  
Therefore, the outer planet must only have $g'+h'$=const, and we may
choose without loss of generality the constant to equal zero.  
Hence, elimination of the nodes implies
$g'=\pi-h$

\subsection{\bf Proof }
Our proof is based on the
triangle equality
$G^2 + G'^2 +2GG'\theta= (H+H')^2$,
which expresses the square of the vertical angular momentum in two 
equivalent ways.
Taking the time derivative of the triangle equality, and noting that
both $G$ and $G'$ satisfy Hamilton's equations, i.e.
$dG/ dt=-{\partial E_{\rm eon}/ \partial g}$ and
${dG'/ dt}=-{\partial E_{\rm eon}/ \partial g'}$,
we find that to leading order in $m$, 
${dG'/ dt} = -{d\over dt}G\theta$, 
i.e.
${dH/ dt} = {\partial E_{\rm eon}/\partial g'}$
which proves that if we replace $g'\rightarrow\pi-h$ in 
$E_{\rm eon}$, we will end up with the correct Hamilton
equation for $H$.

\subsection{\bf  Equations of Motion}

We may re-scale all momenta by an arbitrary
constant without changing the equations of motion 
as long as we rescale the Hamiltonian by the same constant.
Therefore defining
\beqn
J&\equiv&{G\over m\sqrt{G_NM_* a}}\equiv \sqrt{1-e^2} \ \label{eq:gbar} \\
J_z&\equiv& J\cos I\equiv \theta\sqrt{1-e^2} \ , \label{eq:jz2}
\eeqn
the rescaled Hamiltonian is
\beqn
{{\cal H}}(J,g;J_z,h)&\equiv& {E_{\rm eon}\over m\sqrt{G_NM_* a}} \\
&=&-{m'\over M_*}\Omega_*\alpha^3 f(\alpha,e',e,g',g,\theta) \ ,
\label{eq:ham}
\eeqn
where $\Omega_*$ is the orbital angular speed of the test particle around the star.

The  expression for $f$ (defined via eq. [\ref{eq:ee}]) is derived, e.g., in
\cite{Ford00}  to octupole order.
We re-express their Equation (22)
in the limit $m,m'\ll M_*$ as
\beqn
f = {3\over 8}{1\over (1-e^{'2})^{3/2}} (F_{\rm qu}+    {\alpha e'\over 1-e^{'2}}F_{\rm oc}) \ ,
\label{eq:fdef}
\eeqn
where $F_{\rm qu}$ and $F_{\rm oc}$ are displayed above  
(Eqs. \ref{eq:fqu} and \ref{eq:foct}, with
$g\equiv \omega$ and $h\equiv\Omega$)
after correcting the sign of the octupole term
\citep{Ford00error} and
replacing
$g'\rightarrow\pi-h$

The equations of motion are Hamilton's equations for Hamiltonian
(\ref{eq:ham}).  These are displayed explicitly 
in the body of the paper (Eqs. [\ref{eq:eom1}]-[\ref{eq:eom4}])
after
defining the rescaled time
\be
t\equiv {\rm T}\times
{m'\over M_*}\Omega_* 
\alpha^3{3\over 8}{1\over (1-e^{'2})^{3/2}}  \ ,
\label{eq:taudef}
\ee
where $T$ is the true time.

\end{document}